Chapter VI

# Rohingya Refugee Crisis and the State of Insecurity in Bangladesh

Hossain Ahmed Taufiq

The ongoing Rohingya refugee crisis is considered as one of the largest human-made humanitarian disasters of the 21st century. So far, Bangladesh is the largest recipient of these refugees. According to the United Nations Office for the Coordination of Humanitarian Affairs (UN OCHA), approximately 650,000 new entrants have been recorded since the new violence erupted on 25 August 2017 in the Rakhine state of Myanmar.[1] However, such crisis is nothing new in Bangladesh, nor are the security-related challenges new that such an exodus brings with it. Ever since the military came to power in Myanmar (in 1962), Rohingya exodus to neighbouring countries became a recurring incident. The latest mass exodus of Rohingyas from Rakhine state of Myanmar to Bangladesh is the largest of such influxes. Unlike, the previous refugee crisis, the ongoing crisis has wide-ranging security implications on Bangladesh. They are also varied and multifaceted. This crisis erupted at a time when globally we are witnessing an ascendancy of new notions of human security. Exponential growth in religion-centric violent extremism, ethnic cleansing, cybercrime, biological, radiological, chemical and medicinal crime has significantly broadened and deepened the field of security in recent times. Thus, responsibilities for ensuring effective protection have become operationally multilateral.

---

[1] OCHA's latest report indicates that nearly 6,46,000 Rohingyas entered Bangladesh since August 25. See: "Rohingya Refugee Crisis," October 2017. Cited in http://interactive.unocha .org/em ergency/2017_ rohingya/. Accessed on December 11, 2018.

The problem of security regarding the Rohingya refugee issue is complicated by the Islamist insurgency, illicit methamphetamine/yaba drug trafficking, and HIV/AIDS/STI prevalence factors.

The chapter examines the different dimensions of security challenges that the recent spell of Rohingya exodus brings to Bangladesh and the refugees themselves. In order to understand the challenges, firstly the chapter attempts to conceptualise the prominent security frameworks. Secondly, it examines the context and political economy behind the persecution of Rohingyas in the Rakhine state. Thirdly, it explores the political and military aspects of security. Fourthly, it explores the social and economic dimensions. Finally, it examines the environmental impacts of Rohingya crisis in Bangladesh.

**Security and Securitisation**

In this increasingly globalised world, security ranks prominently among the complexities facing the humanity.[2] In the name of security, every day, scores of people are killed, tortured, mutilated, raped, imprisoned, starved, impoverished, displaced, or denied education, health and other basic needs across the globe. The concept of security, however, is extremely saturated. It touches almost all aspects of contemporary societies around the world. It has the power to upset the political, economic and social balance. Today, images of security and insecurity flash across our television screens and the internet almost constantly. Newspaper outlets and radio waves are busy to cover it. All this makes security a fascinating, often deadly, but always an important topic. However, it remains a difficult concept to define even today. Some scholars consider 'security as a beauty: a subjective and elastic term, meaning exactly what the subject in question says it means; neither more nor less.'[3] However, most scholars within political science and international relations work with a definition of security that involves the alleviation of threats to cherished values. As Jarvis argues, the 'perception of a threat is certainly as important as any construction of an objective threat….'[4] Father of modern peace studies Johan Galtung indicated that there are three forms of violent activities that threatens human life- 'direct violence' representing behavior that directly threatens human physical capacity to operate, 'structural violence' representing systematic discrimination against particular groups, finally 'cultural violence' representing the existence of prevalence or prominent social norms that makes direct, structural violence seem "natural".[5] According to noted security expert Barry Buzan, human collectiveness is mainly threatened on five areas – political, military, economic, societal, and environmental.[6] Based on the said conceptual works, the chapter seeks to investigate what security vulnerabilities or threats the Rohingyas are facing while living in Bangladesh as documented and undocumented refugees in the country. Also, it attempts to examine the challenges the recurring Rohingya crisis is bringing to the local economy and society.

---

[2] Barry Buzan, *People, States & Fear: An Agenda for International Security Studies in The Post-Cold War Era* (London: Harvester Wheatsheaf, 1991), p.1.

---

[3] Paul D. Williams, *Security Studies: An Introduction* (Oxon: Routledge, 2008), p.1.
[4] Robert Jervis, *Perception and Misperception in International Politics* (Princeton: Princeton University Press, 1976), p. 6.
[5] Johan Galtung, "Violence, Peace and Peace Research," *Journal of peace research*, pp. 167-191, 1969; Johan Galtung, "Cultural Violence," *Journal of peace research*, pp. 291-305, 1990.
[6] Barry Buzan, *People, State & Fear* (Colchester: Transasia Publishers, 1983).

**Methodology**

To explore the question of insecurity relating to the Rohingya influx, firstly a set of qualitative techniques were employed in Kutupalong and Balukhali camps in Ukhia, Cox's Bazar district. Given the nature of the study, a purposive sampling was used to select the respondents.

| Tools/Techniques | Respondent/s |
|---|---|
| Narrative Inquiry | An injured victim of Tatmadaw's brutal military assault. |
| Key Informant Interviews (KII) | Upazila Nirbahi Officer (UNO) of Ukhia, Program officials of BRAC, Practical Action, and Save the Children, WFP-YPSA Distribution Product Supervision. |

In addition to the narrative inquiry and KIIs, an observation technique was used to investigate the environmental and living conditions of the camps. Secondly, a secondary literature review was carried out. Articles from reputed journal archives, books, book chapters, Governmental and Nongovernmental reports were reviewed. Finally, quantitative data on violence associated with Rohingyas were sought from the Bangladesh Peace Observatory (BPO) platform and various news outlets. A descriptive analysis of these data has been performed, and triangulation with qualitative information has been attempted in order to understand the multifaceted security problems that the Rohingya refugee crisis poses to Bangladesh.

**Context of the Crisis**

Myanmar is located within the corridor of South Asia and South East Asia. It witnessed a historic change when Nobel laureate Aung San Suu Kyi's 'National Leader for Democracy (NLD)' ascended to power through a historic election victory in 2015. Her ascension to power ended a 60 years old direct military rule. Hopes were high that NLD, a party backed by democratic values and principles would act as a guarantor of ensuring the long-term aspiration of peace and prosperity for 50 million Myanmar's people.[7] However, soon it became apparent that due to various constitutional amendments and restrictions, Suu Kyi cannot assume the full and active control of power. She has to share power with the military and retain many of the policies adopted by the previous military-backed 'Union Solidarity and Development Party (USDP)'.[8] Such arrangements effectively rendered the chances of a sustainable solution of the long-standing Rohingya issue in the Rakhine state. In fact, to garner the support of the Myanmarese military, Suu Kyi aligned her policy with the pernicious martial doctrines in the region. The strong backing of China and India reinforced the pseudo-military government to perform an ethnic cleansing against the Rohingyas.[9] There are allegations that many Chinese are operating in the Rakhine state due to their massive investment plans and projects in the region. While asked about Chinese activity, a camp dweller stated that:

---

[7] "The Rise of Rohingya Insurgency and Its Implication on Myanmar and India," 9 September 2017. Cited in https:// moderndiplomacy.eu/201 7/09/19/the-rise-of-rohingya-insurgenc y6- and –its -implication-on-mya nmar-and-india/. Accessed on December 6, 2018.

[8] Azim Ibrahim, *The Rohingyas: Inside Myanmar's Hidden Genocide* (Oxford: Oxford University Press, 2016), p.62.

[9] Lex Rieffel, eds, *Myanmar/Burma: Inside Challenges, Outside Interests* (Washington D.C., Brookings Institution Press, 2010); "Dragon Meets Elephant: China and India's Stakes in Myanmar," 12 October 2017. Cited in http://www.thedailystar. net/opinion/perspective /maya nmar-rohingya-refugee-crisis-dragon-meets-elephant-myanmar-1475020. Accessed on 25 January 2018.

> All those symptoms show that Chinese backing prompted the ongoing actions of the Myanmar government in the Rakhine. Huge numbers of Chinese are operating in the state, and they work closely with the Rakhine Buddhists and the local authorities there.[10]

China's and India's support for Myanmar dates back to late 1980s when the military seized power. Both the Asian powers endeavoured to expand their influence in the reconfigured Myanmar to safeguard their national interests, including multi-billion dollar investments in Myanmar, particularly in the Rakhine state.

China, for instance, has several policy objectives in Myanmar, including access to the Indian Ocean, energy security, border stability, and bilateral economic cooperation. The U.S. control of the adjacent Straits of Malacca is a headache for China, since it is the main passageway of the oil and gas it imports from the Middle East and Central Asia. In 1993, China became a net oil importer.[11] Now, it is one of the largest hydrocarbon consumer countries in the world, with a dependency reaching 64.4 per cent of total oil consumption in 2017 (up from 29 per cent in 2000).[12] This growth is expected to continue.[13] About 80 per cent of China's imported oil passes through the Malacca Straits. Such transshipment makes China highly vulnerable, for example, even to an embargo, let alone conflict.[14] Chinese scholars have long advocated importing oil and gas from the Middle East and Africa through pipelines girding Yunan province to the Kyauk Phyu port.[15] Their two-step plan to avoid the Malacca Strait includes, first, to construct a deep-sea port in Myanmar, where the oil and gas would be brought by ships (in liquified format), and second, transfer the oil and gas through separate pipelines to southwest China. China, therefore, has invested heavily in the oil and gas infrastructure development in Rakhine.[16] A 2.45 billion USD pipeline from Western China to Kyak Phyu is already operational, with the gas pipeline stretching 793km and the oil pipeline 771km.[17] These pipelines will provide a secure route for Beijing's imported Middle East crude oil and gas, and significantly reduces China's reliance on oil and gas supplies passing through the U.S. controlled Strait of Malacca. The oil pipeline can transfer 22 million tons of oil per year which is 5-6 per cent of China's net oil import.[18] Myanmar shares a 2,000km long border with China's Yunan province. For decades, China sought an alternate trade route to Africa, Asia, and the Middle-East. Since Myanmar offers that opportunity, China has invested upwards of 15 billion USD in Myanmar under the sponsorship of its' "The Great International Yunnan Passage."[19] This route comprises of a comprehensive set of road, rail, and air connections through Kunming-Ruili-Muse- Mandalay-Yangon and finally culminates to the Rakhine state.

---

[10] Local Rohingya informant, Kutupalong camp, Ukhia, 24 March 2018. Due to security reasons, he refused to disclose his name.
[11] Jürgen Haacke, *Myanmar's Foreign Policy: Domestic Influences and International Implications* (New York: Routledge, 2016).
[12] ibid.
[13] "China's Oil Import Dependency Deepens,". Washington DC,13 January 132017. Accessed February 22, 2018.Cited in https://oilprice.com/Latest-Energy-News/World-News/Chinas-Oil-Import-Dependency-Deepens.html. Accessed on 22 February 2018.

---

[14] Zhao Hong, "India and China: Rivals or Partners in Southeast Asia?" *Contemporary Southeast Asia*, pp. 121–143, 2007.
[15] "China Moves to Revive its Sway in Myanmar." 28 February 2016. Cited in https://www.wsj.com/articles/china-moves-to-revive-its-sway-in-myanmar-1456697644. Accessed 23 February 2018.
[16] "Dragon meets elephant: China and India's stakes in Myanmar," *op.cit.*
[17] ibid.
[18] ibid.
[19] ibid.

Aung Sun Suu Kyi's 2015 election-win and her ascension to power have brightened the India-Myanmar relationship. India considered it as a golden opportunity to revive the 1991 "Look East" policy, now dubbed as "Act East" to forge economic relations.[20] Since then bilateral relations have improved markedly. Also, India considers Myanmar a key ally in maintaining the security and stability of Northeast India NEI (North East India). The four NEI states— Manipur, Mizoram, Arunachal Pradesh and Nagaland—share a common border of 1,643 km with Myanmar.[21] India considers cooperation with Myanmar a critical stepping stone to limiting the influence of Naga insurgency.[22] Another key reason for New Delhi's new "Act East" policy is to counter China's growing clout in Southeast Asia and the Bay of Bengal. The convergence of all these aspects has impelled New Delhi to invest heavily in Myanmar, setting forth several maritime and land-based Myanmar infrastructural development plans, such as the landmark Kaladan multimodal project, India-Myanmar-Thailand Asian Trilateral Highway, and a road-river-port cargo transport project.[23]

To secure the economic objectives, both China and India extended military cooperation with Myanmar government. Since 2000, both countries have signed more than dozens of deals with Myanmar relating to military hardware sell and training of Tatmadaw.[24]

---

[20] "India's 'Look East' – 'Act East' Policy: Hedging as a Foreign Policy Tool," 5 June 2017. Cited in https://www.fiia.fi/en/publication/indias-look-east-act-east-policy. Accessed on March 21, 2018.

[21] "No Border Dispute between India and Myanmar States Government," 1 August 2018. Cited in https://economictimes.indiatimes.com/news/defence/no-border-dispute-between-india-and-myanmar-states government/article show/65229882.cms. Accessed on December 6, 2018.

[22] "Why Do China, India Back Myanmar over Rohingya crisis?". 18 October 2017. Cited in http://www.scmp.com/week-asia/geopolitics/article/2115839/why-do-china-india-back-myanma r-over-rohingya-crisis. Accessed on 21 March 2018.

[23] ibid.

Despite the strong Chinese and Indian backing, the Rohingya issue has severely weakened Suu Kyi's status in the global community. There are also talks of stripping her off the Nobel Peace Prize. The counter-offensive measures following the insurgency attack conducted by the Myanmar army seriously jeopardised the chances of elevating the long cherished external prospects of Myanmar after a 60-year isolationist policy. International community sympathised with the persecuted Rohingya people and condemned the Suu Kyi's government.

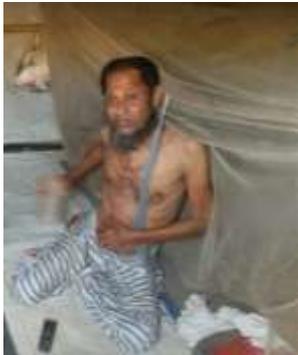

**Name: Abdul Karim**
**Age: 50**
**Location: Kutupalong camp**
**Zone: DD, Camp: 3, Upazilla: Ukhia**

My village was by the seaside. It was a day before Eid-ul Azha when they attacked. At first, in the morning, Myanmar Navy ships fired at the villages at the seaside, then attacked with bombs and rockets from the air. First, they attacked our village. By noon, we fled to the nearby forest and took our cattle and other livestock along with us. We returned to our village when we saw that the firing stopped and there was no one there. However, in the evening, we saw from the distance that a large number of military men approaching towards our village. We took everything once again and started fleeing towards the forest, and they started coming after us and right before we entered the jungle, they started firing at us all at once. They

---

[24] Lex Rieffel, *op.cit.*; "Dragon Meets Elephant: China and India's Stakes in Myanmar," *op.cit.*

> (army) were running after us from the south side and we were trying to escape through the jungles, and when they realised that we were outrunning them, they started firing, and a bullet hit me near my chest. My son got shot in the head near his ear. He instantly fell in front me. He died! He was only 12 years old. (Respondent was silent for some time).
>
> When the bullet hit me, I fell down. People who were fleeing with us carried me to the forest. When we crossed the border and came here in the camp, MSF (Medecins Sans Frontieres, Bangladesh) received us from there. MSF reviewed my condition and referred me to a Cox's Bazar hospital. At the hospital, I had to spend almost three to four months. The bullet hit me (Again he was showing his bullet mark) here but later it went inside my body. They got the bullet out but could not stitch me properly. 'Pus' comes out of my injury every now and then. But, I truly appreciate the overall help that we received from the government of Bangladesh and MSF. I think that is a lot. They kept us for three-four months and provided whatever sort of treatment I needed.
>
> Lot of them died, even the ones that were with us, a lot of them were slaughtered and were carried away in vehicles and a lot of them were captured and lot others were thrown into trenches. But we were lucky enough who were able to escape.
>
> In this shed, me, my two daughters and my wife are now living. We are being provided with the hygiene kits, and so is everyone, some miscellaneous other stuff. We also receive cooking supply such as lentils and rice. Water shortage is a problem in the camp. There is a lack of water sources near my shelter. We have to go very far to fetch water; it is also very crowded. We have to stand in a long queue. If a tube well is installed nearby, it would have been an immense help.

## Political and Militancy Dimension

Barry Buzan's popular security framework indicates that human collectiveness is challenged in two areas – 'political',

which is the stability and system in states and 'military' which ensures the defense of sovereignty and national/ inter-state security. If the two dimensions are considered together, the Rohingya exodus raises several security concerns such as extremism, insurgency, crime, inter-border smuggling, and host/refugee camp security.

**Insurgency and Islamist Militancy**

A growing nexus of secessionist movements and religion-centric extremism in the Arakan region of Myanmar and neighbouring North East Indian states cast a deep shadow in the refugee crisis. Continued denial of citizenship and categorisation of the Rohingyas as illegal Bengali immigrants have contributed to the related problems of insurgency and extremism.[25] In recent times, the Myanmar military – 'Tatmadaw' has opted for a new strategy to oust the so-called illegal Bengali immigrants from the Rakhine territory.[26] In order to force the Rohingyas to leave the state, Tatmadaw is using every conceivable strategy including torture, arson attack, mutilation, rape, killing, and shooting. Not only such actions led to an escalation of hostilities between the ethnic Muslim Rohingyas and Buddhist Rakhines, but also spawned a potent of terror in the region. The extreme punitive measures imposed by the Myanmar government against the so-called illegal immigrants led to the birth of several militant and insurgent groups. One such group is 'Harakah al- Yaqin', notoriously referred in English as the 'Arakan Rohingya Salvation Army (ARSA)', the group responsible for the 25 August (2017) attack.[27] The organisation originated in Saudi Arabia, when Ata Ullah, the leader of the group and his 20 other senior compatriots decided to open the organisation in 2013.[28] The group maintains a strong channel of networks in Saudi Arabia, Pakistan, India, Bangladesh and in many other Muslim countries.[29] ARSA issues fatwa to justify their actions and uses many tactics to attract Rohingyas to join them in their fights against the Tatmadaw and the Myanmar government. In 2016, the group carried out its first coordinated strike, attacking border posts along the border of Bangladesh and Myanmar. On 25 August 2017, there was a coordinated attack on the police check post and military bases in the townships of Maungdaw, Buthidaung and Rathedaung in the Rakhine state of Myanmar. The group claimed responsibility for it.[30]

---

[25] Francis Wade, *Myanmar's Enemy Within: Buddhist Violence and the Making of a Muslim 'Other'* (Chicago: Zed Books, 2017).
[26] Chutima Sidasathian, *Rohingya: The Persecution of a People in Southeast Asia* (California: CreateSpace Independent Publishing Platform, 2012)

In 2004, Bangladesh experienced a coordinated grenade attack allegedly carried out by a religion-centric group on the then opposition party, the Awami league. It left 24 dead and more than 300 injured.[31] The country has suffered several terror attacks since then. Most recent attacks are the Holey Artisan bakery and Sholakia terror incidents. Now, a new Rohingya influx has brought brand-new extremism-related challenges for the country. Earlier studies show that whenever Rohingya influx happened in Bangladesh, it created significant opportunities for militant groups.[32] ARSA is not the only group who are operating in the Rakhine. Over the last three decades, intermittent but pernicious military campaigns carried out by the Myanmar government resulted in forming several armed groups. Groups such as the Rohingya Solidarity Organisation (RSO) and the Arakan Rohingya Islamic Front (ARIF) are operating there since the 1970s.[33] In 1996, RSO and ARIF jointly formed Rohingya National Alliance (RNA).[34] In 1998, ARIF and two other factions of RSO merged and led to the creation of Arakan Rohingya National Organization (ARNO).[35] Other armed groups operating in the border of Myanmar and Bangladesh include the National Unity Party of Arakan (NUPA) and the Arakan Army.[36] Since 1992 and onwards, each new Muslim Rohingya refugee influx and the Refugee camps in Bangladesh have become a golden opportunity for them to recruit people. Today, the number of Rohingya insurgents remain unknown. Earlier estimates suggest that the number is around 1000.³⁷ But, given the number of refugees who have arrived, there is a probability that the number has increased manifolds. The extreme brutality imposed by the Burmese government resulting in anger, grief and vengeance among refugees, and external militant network support further exacerbation of the situation. These groups possible link to Bangladeshi extremist Islamist groups is now one of the greatest sources of concern. A study led by Imtiaz Ahmed in 2010 identifies that several Al Qaeda affiliate groups such as Harakat-ul-Jehad-al Islami (HuJI), Jama'atul Mujahidin, Shahdat-e-Al-Hikma, Hizbut Touheed and Islami Shashontantra Andolon reportedly have bases or operation points at the vicinity of the Rohingya refugee camps.³⁸ The study also confirms that "camps vacated by the Rohingya refugees, and a number of Rohingyas are known to be involved in the smuggling of arms and ammunition in Bangladesh."³⁹

---

[27] "ARSA: Who are the Arakan Rohingya Salvation Army?" 13 September 2017. Cited in https://www.aljazeera .com / news /20 17/09/myanmar-arakan-rohingya-salvation-army-170912060700 394 .html. Accessed on 20 April 2018.
[28] "Myanmar: A New Muslim Insurgency in Rakhine State," 15 December 2016. Cited in https://www.crisisgroup.org/asia/south-east-asia/myanmar/283-myanmar-new-muslim-insurgency-rakhine-state. Accessed on 27 April 2018.
[29] Ibid.
[30] "Myanmar Tips New Crisis after Rakhine State Attacks," 27 August 2017. Cited in https://www.crisisgroup.org/asia/south-east-asia/myanmar/myanmar-tips-new-crisis-after-rakhine-state-attacks. Accessed on 27 April 2018.
[31] "Blasts Hit Bangladesh Party Rally," 22 August 2018. Cited in http://news.bbc.co.uk/2/hi/south_asia/3586384.stm. Accessed on 22 August 2018.

---

[32] Imtiaz Ahmed, "Globalisation, Low Intensity Conflict and Protracted Statelessness/ Refugeehood: The Plight of the Rohingyas", in John Triman, ed., *Maze of fear* (USA: The Newpress, 2004), p. 183; Bertil Linter, Tension Mounts in Arakan State, *Defence Weekly*, 19 October 1991.

[33] "Myanmar and its Rohingya Muslim Insurgency," 7 September 2017. Cited in https://www.csis.org/analysis/myanmar-and-its-rohingya-muslim-insurgency. Accessed on 13 March 2018.

[34] Press Statement of the Rohingya National Alliance (RNA), 26 September 1996.

[35] "Declaration of Arakan Rohingya National Organisations (ARNO)," 13 December 1998. Cited in https://www.rohingya.org/ . Accessed on 14 February 2019.

[36] Imtiaz Ahmed, eds., *The Plight of the Stateless Rohingyas: Responses of the State, Society & the International Community* (Dhaka: University Press Limited, 2010), p. 72.

**Drugs and Arms Smuggling**

Another big security concern for Bangladesh is the illegal trade of drugs and arms via the Bangladesh and Myanmar border. A drug trade and smuggling ring operating in Sittwe and Moungdaw in the Rakhine state and Teknaf of Bangladesh.⁴⁰ Myanmar is the member of the infamous 'Golden Triangle' which refers to three illicit drugs producing countries (others are Thailand and Laos). The country is also located in the proximity of another Narco-producing region the 'Golden Crescent'. Therefore, Myanmar is considered as the biggest hub for both illicit Narco-production and trafficking in the South and South East Asia and also in the world. This is a major source of concern for Bangladesh, as the country is used as a channel for drugs smuggling, particularly Yaba. The Thai word for "crazy medicine", Yaba is primarily manufactured in the Myanmar-Thai border. This is a synthetic stimulant drug mainly methamphetamine in tablet form used for recreational purposes. Other materials used for the drug is caffeine. It comes in a variety of flavours (including grape, vanilla and orange) and colours (commonly in reddish-orange or green). According to Centre for Substance Use Research, University of Maryland (2013), "Today, the United Wa State Army, the largest drug trafficking organization in Burma, is the primary manufacturer of yaba in Southeast Asia; Thailand is the primary market for these tablets."[41]

---

³⁷ ibid, p. 72.
³⁸ Imtiaz Ahmed (2010), *op.cit.*, p. 72.
³⁹ ibid.
⁴⁰ Altsean, *Burma, Report Card: Balancing Act*, March 2003.

Merciless military campaign in the Rakhine made many Rohingyas desperate to cross into Bangladesh. In such circumstances, many of them agree to smuggle illicit drugs in return for safe passage or for meager sum of money.[42] Certain Bangladeshi smuggling lords use them as safe bait. These Rohingya individuals are compelled to go to the border, along with local fisherman to collect the drugs from incoming refugees. There are allegations that some government officials and law enforcement members are also involved in facilitating the drug trade.[43] Smugglers collect the drugs from the fleeing refugees who hide the narco items among their belongings. "It's the perfect disguise—thousands of people coming over the border at once makes it impossible for authorities to track who or what is crossing with them."[44]

---

[41] "Yaba," 2018 Cited in http://www. Cesar .umd .edu /cesar /drugs /yaba.asp. Accessed on 1 May 2018.
[42] "The Unwilling Smugglers," 2018. Cited in http://road sandk ingdoms.com/2018/rohingya-drug-mules/. Accessed on 1 May 2018.
[43] ibid.
[44] ibid.

In 2017, a month after the latest spell Rohingya exodus began, the Bangladesh Border Guard arrested 3 young Rohingya men who were allegedly attempting to smuggle 20,000 pills across the border. [45] On September 2017, Bangladesh police started an investigation against several officers allegedly linked with yaba smuggling to the Cox's Bazar constabulary. Local newspaper reports that the implicated officers had failed to declare 722,000 yaba pills. Authorities suspect that they had sold the drugs for $950,000.[46]

**Table 1**
**Major drugs and arms recovery drives relating to Rohingyas (August 2017- April 2018)**

| Date | Location | Details |
|---|---|---|
| 27 September 2017 | Sitakunda Upazila | Police detained two Rohingya teenagers for attempting to smuggle 1,500 yaba tablets in Sitakunda[47] |
| 28 September 2017 | Naf river estuary | Bangladesh Police arrests three Rohingya men trying to smuggle 8,00,000 meth pills from Myanmar[48] |

---

[45] "3 Rohingya Youths Held with 20,000 Yaba in Teknaf," 30 September 2017. Cited in https://www.thedailystar.net/country/3-rohingya-youths-held-with-20000-yaba-in-teknaf-1469989. Accessed on 17 April 2018.
[46] "The Unwilling Smugglers", *op.cit*.
[47] "Two Rohingya Teenagers Caught Smuggling 1,500 Yaba Tablets in Sitakunda," 27 October 2017. Cited in https://bdnews24.com/bangladesh/2017/09/27/two-rohingyas-teenagers-caught-smuggling-1500-yaba-tablets-in-sitakunda. Accessed on 1 May 2018.
[48] "Bangladesh Police Arrests Three Rohingya Men trying to Smuggle 8 Lakh Meth Pills from Myanmar," 28 September 2017. Cited in https://www.firstpost.com/world/bangladesh-police-arrests-three-rohingya-men-trying-to-smuggle-8-lakh-meth-pills-from-myanmar-4089363.html. Accessed on 27 April 2018.

| Date | Location | Details |
|---|---|---|
| 11-13 October 2017 | Teknaf Upazila | In a gun fight with RAB, two notorious drug smugglers- Drug King Baittya Faruk and Drug Don Didar were killed. Baittya Faruk was killed on Thursday night while Didar was killed on October 12. RAB sources said the RAB-7 teams raided the four listed Yaba smugglers houses at Teknaf on Wednesday last. The team arrested 31 Rohingya people from there and recovered huge yaba.[49] |
| 28 October 2017 | Ukhia Upazila | In a joint operation, Rapid Action Battalion (RAB) and Police arrested three Rohingyas, in the Ukhia camp. They also recovered two small arms, six rounds of bullets, 9,790 pieces of Yaba.[50] |
| 12 November 2017 | Chittagong Port Thana | Department of Narcotics Control (DNC) arrested a Rohingya refugee along with 2000 pieces of Yaba where they also arrested another drug peddler along with 3200 pieces of Yaba from Chittagong city.[51] |

---

[49] "RAB Conducts Special Raids Against Trafficking of Narcotics, Arms," 22 October 2017. Cited in http://the dailynew nation.com/news /151831/rab-conducts-special-raids-against-trafficking-of-narcotics-arms.html. Accessed on 25 April 2018.

[50] "RAB and Police Arrested Three Rohingyas from the Ukhia Camp with Small Arms and Yaba," 30 October 2017. Cited in http://epaper.ittefaq.com.bd/2017/10/30/images/06_110.jpg. Accessed on 24 April 2018.

[51] "DNC Arrested Drug Peddlers Including a Rohingya Refugee with 5,200 Yaba Pieces," 12 December 2017. Cited in

| Date | Location | Details |
|------|----------|---------|
| 22 November 2017 | Cox's Bazar Sadar Upazila | In a raid in Teknaf, , Department of Narcotics Control arrested 5 Rohingyas with 40000 Yaba pieces during a raid.[52] |
| 12 December 2017 | Patiya Upazila | Department of narcotics arrested 2 Rohingya men in Chittagong.[53] |
| 24 December 2017 | Teknaf Upazila | Border Guard Bangladesh (BGB) arrested two Rohingya drug peddlers along with 9 thousand 821 pieces of Yaba in a drive from Teknaf, Cox's Bazar.[54] |
| 6 March 2017 | Cox's Bazar Sadar Upazila | RAB arrested two drug peddlers including a Rohingya with huge Yaba pills.[55] |
| 16 April 2017 | Teknaf Upazilla | Rapid Action Battalion (Rab) members seized 29 lakh yaba tablets.[56] |

http://www.edainikpurbokone.net/content/2017/2017-1212/zoom_view/2m.jpg. Accessed on 24 April 2018.

[52] "In Teknaf, Department of Narcotics Control (DNC) arrested 5 Rohingyas with 40000 Yaba," 24 November 2018. Cited in http://epaper.ittefaq.com.bd/2017/11/24/images/07_100.jpg. 25 April 2018.

[53] "Five, Including Two Rohingyas, Held with Yaba," 13 December 2017. Cited in https://www.thedailystar.net/city/five-including-two-rohingyas-held-yaba-1504228. 27 April 2018.

[54] "Two Rohingyas Captured with Yaba Valuing 30 Lac Takas," 26 December 2017. Cited in http://www. Edainik purbokone.net /index.php?page=1&date=2017-12-26. Accessed on 25 April 2018.

[55] "RAB Retrieved 30,000 Yaba; Arrested Two Including a Rohingya," 7 March 2018. Cited in http://www.edainikpurbokone .net/index. php?page=1&date=2018-03-7. Accessed on 24 April 2018.

[56] "29 Lakh Yaba Tablets Seized in Ctg, Teknaf," 6 April 2017. Cited in https://www.thedailystar.net/country/98-lakh-yaba-tablets-recovered-teknaf-1391878. Accessed on 5 May 2018.

**Employment Issues**

The refugee crisis brought mixed results in the employment sector of Cox's bazar. Locals are enjoying new employment opportunities due to the massive aid and NGO activities. However, the lack of legal status effectively prevents the refugees to seek wage-earning opportunities outside of the camp. Illiteracy is another problem for them. They have no right to own movable or non-movable properties. [57] The government of Bangladesh (GoB) is actively pursuing diplomacy that is spearheaded by the Rohingya repatriation agenda. Under such circumstances, the government is resistant to any future pressure of naturalisation of the refugees. Thus, GoB has opted for a strategy to deter refugees' access to labour markets. Indeed, NGOs and development partners attempted to provide a number of livelihood training to the refugees to start up business. But as authorities later halted or suspended any such training activities, they offered limited prospects.

On the contrary, the sudden influx of refugees brought new opportunities for the local people. Local people, especially the youth took advantage of the language they speak. The language they speak, and the language refugees speak are similar in dialect. It is easier for the locals to assimilate with the Rohingyas. As a result, many local students who previously had no work quickly grabbed the opportunity to work for development partners in various capacities. With only HSC degree they can apply for the positions.[58] Job switching is also becoming common among the daily wage earners. Many of them left their old hard work and joined the NGOs and international agencies which offer attractive remuneration. This created a crisis in the labour market as farmers now struggle to find suitable labourers to work for their firms.

---

[57] Imtiaz Ahmed (2010), *op.cit.,* p. 79.
[58] "Implications of the Rohingya Crisis for Bangladesh," 11 November 2017. Cited in http://cpd.org.bd/wp-content/uploads/2018/0 1/presentation-Implications-of-the-RohingyaCrisis-for-Bangladesh.pdf. Accessed on 10 December 2017.

**Camps related Security Issues**

In the camp area, refugee's freedom of movement is severely restricted. Leaving the camp area without authorised permission and escort can result in arrest and detention. Such restriction may be necessary for the authorities to ensure order and security, but it also bars refugees' access to education, life skills opportunities and other public services. They remain locked in aid dependence and idleness and cannot reduce their economic and psycho-social stresses. Such stressors increase the chances of their pursuit of harmful coping mechanisms, such as child or forced marriage, survival sex, or exploitation and gender-based violence (GBV). Restrictions also negatively impact the intracommunity cohesion and peaceful cohabitation with the host Bangladeshi communities.

According to Bangladesh constitution refugees are not allowed to receive citizenship or any legal status, thus effectively rendering them unable to access civil administration services and justice. They cannot have births, deaths or marriages registered. Any formal certification for education is disallowed. Their access to formal medicolegal reports documenting criminalising acts in Bangladesh is denied, including reports of sexual harassments, rape and domestic violence. The absence of legal status also leaves them exposed to unlawful detention. The detention facilities are not often gender-segregated, which is a grave concern for girls and women in conflict with the law.⁵⁹ A sluggish establishment of civilian governance mechanism and civil administration combined with lack of rule of law in the refugee camps force the refugees to rely on informal camp-based mechanisms for dispute resolution and civil service. There are allegations that these informal systems are abusive in nature. Overall, the protection system in the camps are almost dysfunctional, and refugees can only rely on aid and luck. Furthermore, lack of alternative opportunities makes Rohingya men, women and children vulnerable to debt bondage, cheap and exploitative labour, child labour, forced labour, human trafficking, survival sex, and exploitation.⁶⁰

**Crime and Violence**

Hosting a large number of refugees is not a pleasant experience for a society. Recurring incident such as the Rohingya crisis is no exception. The region where they have taken shelter is one of the poorest places of Bangladesh. Mountainous location, remoteness, and lack of cultivable lands are the reasons behind the abject poverty there. Besides, an extra pressure of incoming refugees created a huge tension among the local Bangladeshi inhabitants. Recent rise in violent and criminal incidents relating to the Rohingyas are reflective of the tension. Drug peddling, internal feuds, mob violence are becoming prominent since August 2017. According to Bangladesh Peace Observatory (2018), a total of 42 major criminal incidents took place since the recent crisis started (from August 2017 to April

---

⁵⁹ "JRP for Rohingya Humanitarian Crisis," 16 March 2018. Cited in https://reliefweb.int/sites/reliefweb.int/files/resources/    JRP%2 0for%20Rohingya%20Humanitarian%20Crisis%20-%20FOR%2 0DISTRIBUTION.PDF. Accessed on 2 May 2018.
⁶⁰ ibid.

2018).[61] In these criminal offences, both Bangladeshi nationals and Rohingyas have fallen as victims. For example, in January 2018, a group of unidentified miscreants killed a Rohingya Muazzin in the Balukhali camp, Cox's bazar.[62] Officials claim that this was the third murder in a series of killing took place against the people who openly favours repatriation. In another incident in October 2017, a Rohingya man attacked a Bangladeshi man for allegedly having an illicit relationship with his female relative.[63] Just sometime after the incident, based on robbery allegations, Rohingya inmates attacked the nearby Bengali community and injured five Bangladeshi workers in the Balukhali camp.[64] In January 2018, a Rohingya man stabbed another to death at Kutupalong based on old enmity.[65] Most recently, Rohingyas beat up four Chakma in the Ukhia upazilla.[66] Injured indigenous group members were later taken into the nearby hospital for treatment. The incident took place when Rohingyas wanted to collect firewood by violating a betel leaf garden and faced protest from the Chakma owners. Illegal chopping of trees is another source of violence frequently committed by the refugees. On January 2018, local Forest Department chased several Rohingyas away in an eviction drive where ten Rohingyas and eight forest officials received injuries. At least 200 temporary sheds were demolished built by them in the reserved forest areas in Kutupalang, Ukhia.[67] Similar incidence of hacking, killing, illegal activities are taking place in a regular interval. Bangladesh Police, Rapid Action Battalion (RAB), and Border Guard Bangladesh (BGB) have been deployed to prevent such incidents. So far, the law enforcer agencies are actively carrying out their duties. But, mountainous terrain, dense forest, and fragile border severely hinder their efforts.

---

[61] Bangladesh Peace Observatory, *Violence data- Rohingya (*Dhaka: Centre for Genocide Studies, 2018).
[62] CGS, *CGS Peace Report, Volume 2, Issue 1*. (Dhaka: Centre for Genocide Studies, 2018).
[63] "A Rohingya Man Attacked a Bangladeshi Man for Allegedly having an Illicit Relationship with His Female Relative," 29 October 2017. Cited in http://epaper.ittefaq. com.bd/2017/ 10/29 / images/23_105.jpg. Accessed on 17 April 2018.
[64] "5 injured in Attack by 'Rohingya Miscreants'," 28 October 2017. Cited in https://www.thedailystar.net/rohingya-crisis/4-injured-attack-rohingyas-refugees-myanmar-1482943. Accessed on 17 April 2018.
[65] "Rohingya Stabbed to Death at Camp," 14 January 2018. Cited in https://www.thedailystar.net/country/rohingya-stabbed-death-camp-1519519. Accessed on 17 April 2018.
[66] "Rohingyas Beat up Four Chakmas," 18 March 2018. Cited in http://epaper.ittefaq.com.bd/2018/03/18/images/04_108.jpg. Accessed on 15 April 2018.

**Social and Economic Dimension of Security**

Economic security revolves around the resources, markets and finance necessary to maintain welfare and state power. While, societal security centered on the preservation of language, culture, religious and national identity and custom. The economic and social security issues concerning the refugee crisis are illicit trade, inflation and commodity price, Gender based violence prostitution and health related threats.

**Supply of Aid and Illegal Trade**

Both UNHCR and WFP are providing sufficient amount of rations to the refugees at the Kutupalong and Balukhali camps. Activities of WFP indicated that there is an ample supply of rations for the refugees. While asked about the distribution a WFP worker informed us that:

> We are giving people food support on behalf of WFP… Three items. One is rice, we're giving each

---

[67] "10 New Rohingya Refugees Injured in Forest Department's Eviction Drives," 6 January 2017. Cited in http://www.newagebd. net/article/6358/10-new-rohingya-refugees-injured-in-forest-departments- eviction -drives. Accessed on 6 December 2018.

> household almost 20 kg rice, 30 packets in each of them. Then there is oil. We generally give them 3 litres of oil, but we are providing them with 2 litres due to a shortage in the warehouse, but we usually provide them with 3 litres. And we give them 9 kg Lentils. That is based as per categories, the supply information is available. The category is based on three criteria, the people who are under our camp are divided into those three criteria and are given provisions.[68]

However, the practice among the Rohingyas about selling part of their rations received from the UNHCR or WFP in the local markets is reported by several newspapers and aid agencies. During the field visits, a local Bengali youth who is working with several aid agencies for some time now informed that:

> On several occasions, aid agencies and law enforcement members found that Rohingya refugees are selling their rations such as lentils, rice and medical kits outside of the camps in exchange for cash and other necessary things. The authorities prohibit selling of the aid items received from the development agencies. But, violations of the ban are not uncommon, and there is a syndicate operating here who help or sometimes coerce the refugees to sell the items.[69]

Earlier field inquiries pointed out the fact that Rohingyas believe if they have a large family staying in the same camp, then they can claim larger quantity of rations.[70]

---

[68] Liton, WFP-YPSA Distribution Product Supervisor, Kutupalong Camp, 24 March 2018.
[69] Tashdid, Young Aid Worker, Kutupalong Camp, Ukhia, 24 March 2018.
[70] Imtiaz Ahmed (2010), *op.cit*.

Afterwards, they can sell the items in to the local market. The same study showed that there is a MOU between Aid agencies and the government to restrict the trading of the aid items.[71] But authorities often overlook such arrangements from a humanitarian stand point. Lentils and rice cannot fulfill all their nutritional demands. They sell their rations because with the money or in exchange they can get some protein items such as meat, chicken and all other necessary ingredients.

**Images 1 & 2: General food distribution, Kutupalong Refugee camp in Cox's Bazar Area (pictures taken from a WFP food warehouse on 25 April 2018)**

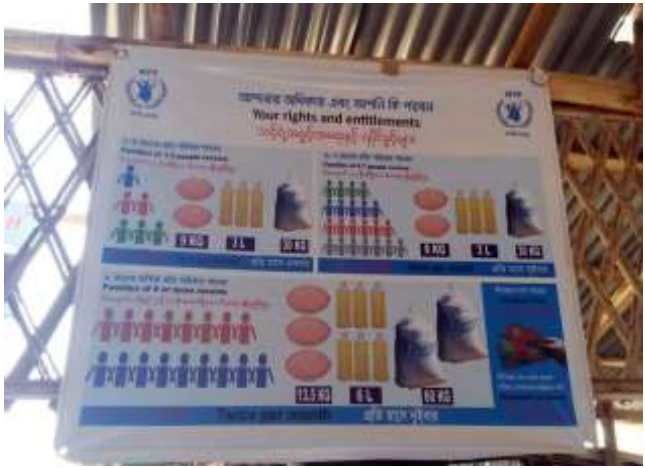

---

[71] ibid.

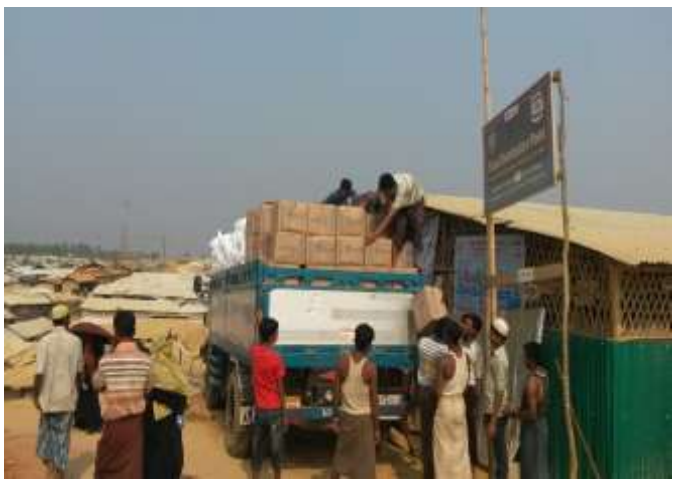

## Inflation and Rising Price Levels

Food and commodity price increase is another source of concern in the Ukhia and Cox's bazar area. Some commodity price at the near bazars skyrocketed. For instance, right before the new influx started, 1kg potato cost 20 takas ($0.24); it has since tripled in price to 60 or 70 taka. The price of fish has also been doubled.[72] There is a growing consternation among the locals as the recent influx threatens to push the land prices in the region.

## Prostitution and STI/RTI Risks

As Rohingya women struggle to access the very basic needs in the overcrowded camp, many of them are compelled to engage in prostitution and survival sex. However, the number of women and girls engaged in such activities is very difficult to determine. During the field visit, officials of BRAC and Practical Actions- two prominent NGOs informed that: "collecting such information is extremely risky, it can seriously agitate the camp dwellers, most of the NGO and aid agency field workers typically avoid such data practices." Even the UN agencies do not maintain any figures on the number of sex workers active in the camps. In an interview with Reuters news agency, Saba Zariv, an expert on gender-based violence at the U.N.'s population agency UNFPA told journalists, "It's hard to come by numbers and we don't collect data on how many sex workers are in the camps."[73] *Therefore,* the only accounts remain are the case studies carried out by different news agencies. Reuters report claim that at least 500 Rohingya sex workers live in Kutupalong.[74] In most of the cases, these girls or women manage to go outside of the camp with the help of locals and meet their customers. Such growing prevalence of prostitution combined with yaba drug abuse is seriously increasing the chance of a rise in HIV/ AIDS and other STI/RTI in the zone. A recent report suggests that at least 62 Rohingyas have HIV/AIDS.[75] These are only documented cases. The actual number could be higher than this. According to Shaheen Md Abdur Rahman Chowdhury, residential medical officer of Cox's Bazar Sadar Hospital, "In the field level, we don't have any mechanism to detect HIV patients. We refer those who have symptoms of HIV to the district hospital for confirmation."[76] In the camp, the hygiene practices are not decent, and many of them have 'risky sexual behaviours'.[77] Apparently, the number of HIV positive is not high. But, in congested camps such as Balukhali and Kutupalong, the risk of spreading of the infection is very high.

---

[72] "Rohingya Refugees Test Bangladeshi Welcome as Prices Rise and Repatriation Stalls," 28 February 2018. Cited in https://www.reuters.com/article/us-myanmar-rohingya-banglade sh-tensions/rohingya-refugees-test-bangladeshi-welcome-as-pric es-rise-and-repatriation-stalls-idUSKCN1GC08Y. Accessed on 4 May 2018.

---

[73] "Clandestine Sex Industry Booms in Rohingya Refugee Camps," 23 October 2017. Cited in https://news. Trust.org/item /20171023230819-d6gas/. Accessed on 1 April 2018.
[74] ibid.
[75] "At Least 62 Rohingya Refugees have HIV/AIDS, Bangladesh Officials say," 13 October 2017. Cited in https:// www.Rfa org /english/news/myanmar/refugees-hiv-11132017155619.html. Accessed on 4 April 2018.

## Environmental Dimension of Security

Buzan's frameworks final pillar concerns with the maintenance of the biosphere and ecosystem. The major issues regarding the environmental security include deforestation, environmental pollution, living condition and loss of habitat.

## Deforestation, Environmental Pollution and Loss of Habitat

In order to accommodate the new Rohingyas, government has allowed to clear the forest in the Kutupalong, Balukhali, and other adjacent areas. There are claims that Rohingyas are cutting the trees and hills indiscriminately here and there, causing significant distortion to the nearby environments. Fuel for cooking is an issue, and since 2004 UNHCR no longer provides firewood, these may have led to their indiscriminate chopping of trees.[78] Already, there are two existing camps for the previously displaced Rohingya in the 'Reserve Forests' of the Department of Forestry.[79] Cleaning of reserve forests for camp purpose resulted in the displacement of local people who used to graze the lands. The recent decision to another round of clearing of forests further endangers the already vulnerable ecosystem in the

---

[76] ibid.
[77] ibid.
[78] Imtiaz Ahmed (2010), *op.cit*.
[79] ibid, p. 83.

Cox's Bazar area. In an interview during the field data collection, the Upazila Nirbahi Officer (UNO) of Ukhia stated that:

> Actually, it is the environment which is suffering the most. 3,000 acres of land has already been allocated before, but it is now raised to 5,000 acres. The entire land was more or less dense forests and hill tracts. So, now the entire forest had to be cut down to accommodate the refugees here. Even on top of the mountains, there are no trees left. We are already witnessing the impact of this deforestation. Now, it is really cold during winter, colder than the earlier winters. And because of the trees being chopped off from the mountains, the risks of landslides have increased significantly. This zone is a tourist zone. The winter here was great, but now we are witnessing a change in that climate pattern. And we are predicting that the coming summer will bring extreme heat. The government is working to tackle this, through various reforestation projects, etc.[80]

According to Centre for Policy Dialogue (2017), the total forest area in the Cox's bazar is around 2,092,016 acres. However, due to recent Rohingya influx, the initial loss of the forest area stands at 3,500 acres, equivalent to 1.67% in Cox's bazar forest area, and 0.05% loss in total national forest area.[81] Another 2000 acres of forest awaits the same fate.[82] The estimated value of the forest land occupied by the Rohingya is worth BDT 5 billion.[83]

---

[80] Md Nikaruzzaman, UNO, Ukhia, 24 April 2018.
[81] "Implications of the Rohingya Crisis for Bangladesh," *op.cit.*
[82] "Shrinking Elephant Habitat: Deforestation Largely Blamed," 18 December 2017. Cited in https://www.thedailystar.net/back page/shrinking-elephant-habitat-deforestation-largely-blamed-1506514. Accessed on 3 May 2018.
[83] "Implications of the Rohingya Crisis for Bangladesh," *op.cit.*

**Figure 3**
**Deforestation rate in the camp areas (Sources: Department of Forestry, GoB)**[84]

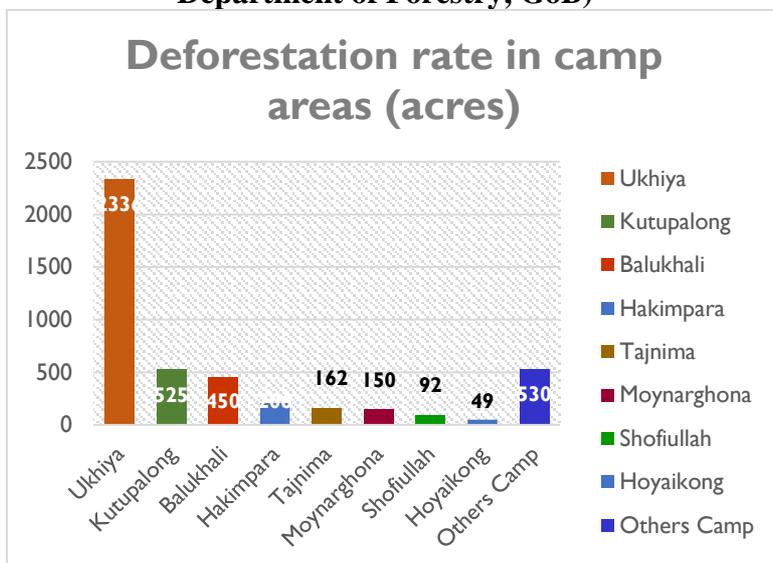

The Rohingya influx in the Cox bazar region and subsequent deforestation have dealt a double blow to the natural ecosystem and habitat. Species living in close proximity to the area are slowly moving out or getting killed or extinct. For example, the wild elephant population became seriously vulnerable for the recent deforestation. International Union for Conservation of Nature (IUCN) claims that "out of eight of 12 elephant corridors in the Bangladesh's nine reserve forests are in the distrcit of Cox's Bazar."[85] The report further claims that "of the 300 elephants in nine reserve forests, 112 have been spotted in Cox's Bazar alone."[86] Acute shortage of food and loss of habitat forced the elephants to venture out for alternative sources, leading to clash with humans. According to a report published by Dhaka Tribune (2018), at least five elephants died since the latest spell of Rohingya influx, which started in August, and at least three of them got killed from electrocution and land-mine related injuries.[87] The Myanmar army has started to install land mines and barbed wires in the border region, which disallow the transboundary migration of the giant mammals. Obstruction of passage and destruction of habitat made the mammals such as elephants to raid the neighboring crop fields. Such invasion are leading to inevitable clash between human and the wild giants.[88] At least twelve Rohingyas have been killed in 2017 by wild elephants whose habitat has been consumed by the recent establishment of Kutupalong refugee camp.[89]

---

[84] ibid.
[85] "Shrinking Elephant Habitat: Deforestation largely blamed," *op.cit*.
[86] ibid.

---

[87] "Rohingya Influx Deals Blow to Bangladesh's Wild Elephant Population," 9 February 2018. Cited in https://www.dhakatribune.co m /opinion/special/2018/02/09/rohingya-influx-deals-blow-bangladeshs-wild-elephant-population/. Accessed on 9 February 2018.

[88] "Elephants Trample 10 Rohingya Refugees to Death in Search for Food," 7 March 2018. Cited in http://www.scmp.com/news /asia/southeast-asia/ article /213 6035 /hungry-elephants-kill-10-rohingya-refugees-bangladesh. Accessed on 2 May 2018.

[89] "Smiles and Slapstick as Rohingya Refugees Learn to Corral Elephants," 19 April 2018. Cited in https://www.iucn.org/news /asia/201804/smiles-and-slapstick-rohingya-refugees-learn-corral-elephants. Accessed on 2 May 2018.

**Figure 4**
**Facts on elephant deaths since the Rohingya influx of 2017 (Source: Dhaka Tribune, 9 February 2018)**[90]

**ELEPHANT DEATHS (NOV 21-JAN 22)**

- An elephant was found dead on the night of January 22 at Ghonarpara in Chakaria, Cox's Bazar. Locals claimed it was shot dead because it came to civilization in search of food
- But the Forest Department claim the mammal died of natural causes after falling sick
- On November 25, an elephant and her calf were electrocuted after getting entangled in a live wire placed by a farmer near Barahatia forest under Chittagong South Division
- On November 25, another female elephant died at Raikhali area in Kaptai upazila of Rangamati. Luckily, her baby escaped unscathed
- The mother elephant is believed to have died after eating poisoned bait set by the farmers to deter wild elephants from raiding crop fields during the pre-harvest period
- On November 21, a wild elephant died after stepping on a landmine allegedly laid by Myanmar security forces in the no man's land along Naikhongchhari border in Bandarban

Environment pollution around the refugee camps is a common scenery at Kutupalong and Balukhali. Proper and adequate sewerage system in the refugee camps are almost nonexistent. Therefore, contamination of water is a grave concern in the camp. All sewerage and fecal sludge are directed to the Naf river.[91] It is seriously contaminating the river water. The prevalence of cholera and diarrhea are quite common in the camps. Another problem is the lack of regular waste management. Dwellers use various materials and litter them here and there. Particularly the sudden influx of refugees greatly increased the consumption and use of humanitarian goods. Piles of left over from these goods such as plastics, packaging of cooking appliance, food or relief assistances made the camps very dirty and clumsy.

---

[90] "Rohingya Influx Deals Blow to Bangladesh's Wild Elephant Population," *op.cit*.
[91] Imtiaz Ahmed (2010), *op.cit*.

Rohingya Refugee Crisis and Insecurity in Bangladesh● 177**Figure 5**
**Pollution at the camp is very high (picture taken from Kutupalong camp on 24 March 2018)**

**Living Conditions**

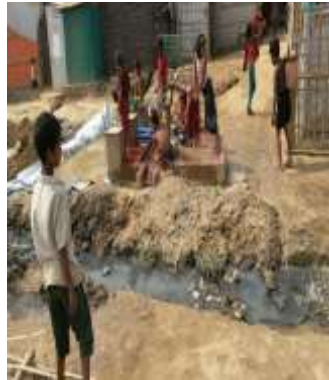

Currently, most of the new refugees are using makeshift shelters constructed with the help of the UNHCR. These makeshift shelters are temporary in nature and grossly below the human living standards. They are overcrowded and extremely vulnerable to the seasonal climatic fluctuations. Flooding and landslides, especially during the rains, monsoon and cyclone season, expose refugees to high risks. Such poor quality along with lack of privacy critically jeopardizes the physical security and psychological well-being of the women, children, old and refugees with disabilities. Overall environment of the camps is very disorganized and dirty. Access to adequate water and sanitation is another problem. Rohingyas are now mainly placed in two types of locations- camp like settings and host communities. [92] Following table presents some facts provided by the International Organization for Migration (IOM) on WASH.

---

[92] "Needs and Population Monitoring (NPM): Site Assessment: Round 8," 5 February 2018. Cited in https://www.humanitarianresponse.info/sites/www.humanitarianresponse.info/files/assessments/npm-r8-sa-report_2018-02-05.pdf. Accessed on 11 October 2018.

**Table 2**
**Access to Water, Sanitation and Hygiene (WASH)**[93]

| Water Sources | Tube wells/handpump are the most common sources of water for 90% household locations. Of these, 11% are in the camp like settings and 79% in the host communities. |
|---|---|
| Water Needs | At least in 7 locations, it is reported that refugees do not do not have access to water at all. 6 of those locations were in the camp like settings. It was reported that only 9% of the locations have limited access to clean water and even then, they can use it for basic needs. |
| Access to Bathing facilities | Some people can bath in about 15% of the locations and around 2% of the locations they do not have any bathing facility. Around 36% of the locations half of the population has access to bathing facilities. They also reported that in around 33% locations most people have access and in around 14% almost all have access to bathing facilities. |
| Access to latrines | Around 14% of the locations everybody has access to latrines, around 36% of the locations most have access, and around 35% of the locations around half have access to latrines. In around 14% of the locations some have access, and in around 1% of the locations, it was reported that no one has access to sanitary latrines. |

---

[93] ibid, pp. 6-7.

Rohingya Refugee Crisis and Insecurity in Bangladesh● 179

**Figure 6**
**Percentages of locations by settlement type and access to bathing facilities (Source: IOM)**[94]

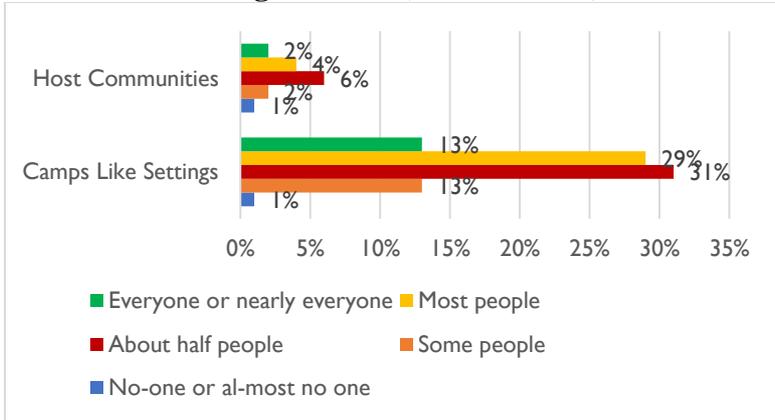

**Figure 7**
**Percentages of locations by settlement type and access to latrine (Source: IOM)**[95]

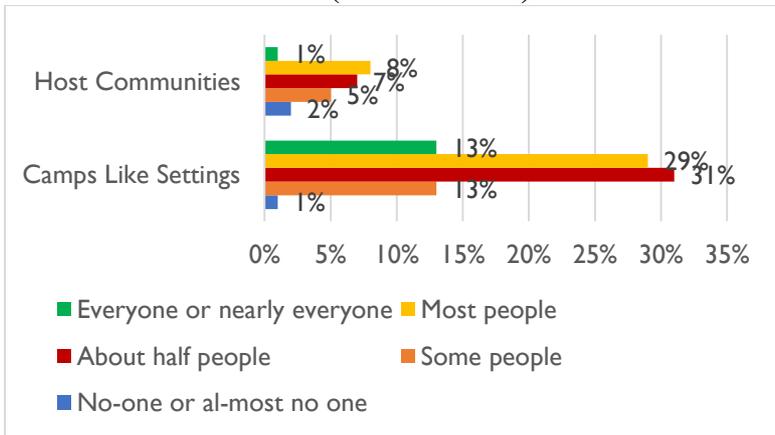

---

[94] ibid
[95] ibid.

## Conclusion

Viewed from the holistic perspective, the present Rohingya refugee issue and security have important connections. The recent spell of refugeehood and its' impact have been discussed from five major security dimensions- military, political, economic, societal, and environmental. Though, it has repeatedly been stated that security is an issue mainly to the host government, other parties such as the refugees themselves, local population, country of origin, sponsors and development organisations face the ripple effects of this crisis. Presently, two different bodies are responsible for ensuring security in the refugee camps. One is the host country government, and the other is the development partners such as United Nations with its various agencies and bodies and NGOs. The primary responsibility lies with the Bangladesh government. They are responsible for ensuring the physical safety of the refugees and maintaining security throughout the refugee camps and settlements. The international community, however, has identified that the host government needs the outside help to fulfil its responsibility properly. The role of the host government, therefore, is very critical. It holds power to determine the fate of the refugees. The role of Rohingya is also essential to keep the security in the camps and settlements. Ultimately. it is the Bangladeshi people who are hosting them during their grim times.